\documentclass[fleqn,usenatbib]{mnras}

\usepackage{newtxtext,newtxmath}
\usepackage[flushleft]{threeparttable}
\usepackage[T1]{fontenc}

\DeclareRobustCommand{\VAN}[3]{#2}
\let\VANthebibliography\thebibliography
\def\thebibliography{\DeclareRobustCommand{\VAN}[3]{##3}\VANthebibliography}

\usepackage{graphicx}	% Including figure files
\usepackage{amsmath}	% Advanced maths commands
\usepackage[justification=centering]{caption}
\usepackage{graphicx}
 \usepackage{setspace}
\usepackage{textcomp}     % access \textquotesingle
\usepackage{url}
\usepackage[]{amsmath}
\usepackage{amsmath}
\usepackage{nccmath}
\usepackage{lineno}
\usepackage{siunitx}
\title[Keck, Gemini, P200 Neptunian Trojan photometry]{Keck, Gemini, and Palomar 200-inch visible photometry of red and very-red Neptunian Trojans}

\author[B. T. Bolin et al.]{
B. T. Bolin,$^{1,2,3}$\thanks{NASA Postdoctoral Program Fellow}$^{,}$\thanks{E-mail: bryce.bolin@nasa.gov (BTB)}
C. Fremling,$^{2,4}$
A. Morbidelli,$^{5,6}$
K. S. Noll,$^{1}$
J. van Roestel,$^{2,7}$
\newauthor
 \;E. K. Deibert,$^{8}$
M. Delbo,$^{5}$
G. Gimeno,$^{8}$
J.-E. Heo,$^{8}$
C. M. Lisse,$^{9}$
T. Seccull,$^{10}$
and H. Suh$^{10}$
\\
$^{1}$Goddard Space Flight Center, 8800 Greenbelt Road, Greenbelt, MD 20771, USA,\\
$^{2}$Division of Physics, Mathematics and Astronomy, California Institute of Technology, Pasadena, CA 91125, USA,\\
$^{3}$Infrared Processing and Analysis Center, California Institute of Technology, Pasadena, CA 91125, USA,\\
$^{4}$Caltech Optical Observatories, California Institute of Technology, Pasadena, CA 91125, USA,\\
$^{5}$Laboratoire Lagrange, UMR7293, Universit\'{e} de Nice Sophia-Antipolis, CNRS, Observatoire de la C\^{o}te d'Azur, Nice, France,\\
$^{6}$Coll\`{e}ge de France, 75231 Paris, France,\\
$^{7}$Anton Pannekoek Institute for Astronomy, University of Amsterdam, 1090 GE Amsterdam, The Netherlands,\\
$^{8}$Gemini Observatory/NSF's National Optical-Infrared Astronomy Research Laboratory, Casilla 603, La Serena, Chile,\\
$^{9}$Johns Hopkins University Applied Physics Laboratory, 11100 Johns Hopkins Rd, Laurel, MD 20723, USA,\\
$^{10}$Gemini Observatory/NSF's NOIRLab, 670 N. A'ohoku Place, Hilo, Hawaii, 96720, USA
}
\date{Accepted XXX. Received YYY; in original form ZZZ}

\pubyear{2022}

\begin{document}
\label{firstpage}
\pagerange{\pageref{firstpage}--\pageref{lastpage}}
\maketitle

\begin{abstract}
Neptunian Trojans (NTs), trans-Neptunian objects in 1:1 mean-motion resonance with Neptune, are generally thought to have been captured from the original trans-Neptunian protoplanetary disk into co-orbital resonance with the ice giant during its outward migration. It is possible, therefore, that the colour distribution of NTs is a constraint on the location of any colour transition zones that may have been present in the disk. In support of this possible test, we obtained $g$, $r$, and $i$-band observations of 18 NTs, more than doubling the sample of NTs with known visible colours to 31 objects. Out of the combined sample, we found $\approx$4 objects with $g$-$i$ colours of $>$1.2 mags placing them in the very red (VR) category as typically defined. We find, without taking observational selection effects into account, that the NT $g$-$i$ colour distribution is statistically distinct from other trans-Neptunian dynamical classes. The optical colours of Jovian Trojans and NTs are shown to be less similar than previously claimed with additional VR NTs. The presence of VR objects among the NTs may suggest that the location of the red to VR colour transition zone in the protoplanetary disk was interior to 30-35 au.
\end{abstract}


%%%%%%%%%%%%%%%%%%%%%%%%%%%%%%%%%%%%%%%%%%%%%%%%%%

%%%%%%%%%%%%%%%%% BODY OF PAPER %%%%%%%%%%%%%%%%%%

\section{Introduction}

Understanding how the composition of the Solar System's protoplanetary disc varied as a function of the heliocentric distance has always been a key problem of planetary science with implications on the formation of planetesimals, planets, meteorites, and the delivery of organics, water and prebiotic materials to planets \citep[][]{Williams2011}. While it is now understood that the contemporary asteroid main belt consists of objects that originally accreted in the terrestrial planet and Jupiter formation regions, as well as objects that were implanted from the primordial Kuiper belt \citep[][]{DeMeo2014a}, the compositional structure of the original trans-Neptunian disc (TND) is not yet understood.

It has been theoretically demonstrated that the original configuration of the TND was a low-inclination formation, starting from 23 au with a drop in density past 30 au \citep[][]{Morbidelli2020KBO}. The TND may have had a colour gradient of trans-Neptunian objects (TNOs) increasing in redness with heliocentric distance due to the sublimation of surface volatiles such as ammonia, methanol and hydrogen sulfide \citep[][]{Brown2011,Wong2016, Schwamb2019}. The present-day TND consists of the Hot Classical objects (HCs), comprised of bodies that may have formed within 30 au but were scattered outwards by the migration of Neptune, and the Cold Classical objects (CCs), comprised of bodies that probably formed outside of 30 au and had much-reduced interactions with Neptune \citep[][]{Morbidelli2020KBO}. The resonant population consists of TNOs that are in mean motion resonances with Neptune located at $>$30 au from the Sun such as the 5:4, 4:3 and 5:3 mean motion resonances at $\sim$34.7 au, $\sim$36.2 au and $\sim$42.4 au. The ratio p:q denotes the resonance of p orbital periods of Neptune to q periods of the TNO \citep[][]{Gladman2008}. Scattered disc objects are TNOs that are on orbits which are currently scattering off Neptune such that their semi-major axes, $a$ change by more than 1.5 au in 10 myrs \citep[][]{Morbidelli2004SDO}. In addition to the HCs, CCs, resonant objects and scattered disc objects, the Neptunian Trojans (NTs) located at $\approx$30 au in the Sun-Neptune L4 and L5 Lagrange points \citep[][]{Sheppard2006}, are hypothesized to have been captured from the TND into co-orbital resonances with Neptune during its outward migration \citep[][]{Gomes2016}.

The colours of TNOs are known to be bimodally distributed between ``red'' (R) and `very-red'' (VR) object colours \citep[][]{Hainaut2012} where R objects are defined as having an optical spectral slope of $\lesssim$20 $\%$ / 100 nm corresponding to a $g-i$ colour index of $<$1.2 \citep[][]{Sheppard2012} in the SDSS $g$and $i$ bandpasses \citep[][]{Fukugita1996}. The ``very-red'' (VR) category of TNOs are defined as having an optical spectral slope $\gtrsim$20 $\%$ / 100 nm corresponding to a $g-i$ colour index of $>$1.2 \citep[][]{Wong2017a}. The HCs are a more equal mixture of ``red'' (R) and `very-red'' (VR) objects while the CCs have a higher ratio in the number of VR objects to the number of R objects \citep[][]{Trujillo2002}. One of the explanations for the colour dichotomy between R and VR objects is that the original TND had a colour transition boundary from R to VR objects occurring in the primordial disc between $\approx$30 and $\approx$40 au \citep[][]{Nesvorny2020}.

Out of 32 known L4 and L5 Lagrange point NTs \citep[e.g.,][]{Sheppard2006,Parker2013,Bernardinelli2022}, 16 have optical colours which cover a wide range in optical slope with the majority having an optical spectral slope of $<$20 $\%$ / 100 nm or $g-i$ $<$ 1.2 \cite[][]{Jewitt2018}. Presently, only one NT is known to have colours that place it in the VR category, 2013 VX$_{30}$ with $g-i$ = 1.52 $\pm$ 0.06 \citet[][]{Lin2019}. The dearth of VR category objects is surprising because the NTs were captured at a similar heliocentric distance as HCs, but HCs have a higher VR to R colour ratio. This suggests that the transition boundary between R and VR objects was actually much further out from where the NTs were captured, more than 30 au from the Sun and possibly as far out as 40 au near the formation region of the CCs \citep[][]{Nesvorny2020}. 

In this work we expand on the previous work available on the visible colours of NTs, with observations of 18 objects, 15 of which are new, which increases the number of NTs with known visible colours to 31. 

\section{Observations}
We obtained optical $g$, $r$/R, and $i$/I photometry of 18 NTs with the Hale 5.1 m telescope (P200 hereafter) at Palomar Observatory, the Gemini North 8.1 m telescope (Gemini N hereafter), and the Keck I 10 m telescope at Maunakea Observatory, and the Gemini South 8.1m telescope (Gemini S hereafter) at Cerro Pach\'{o}n. Observations of our 18 NT targets were divided between the P200, Keck I, Gemini N, and Gemini S during 2020-2022. Five NTs were observed with the P200 using the Wafer-Scale Imager for Prime focus (WaSP) instrument \citep[][]{Nikzad2017}. Three NTs were observed with Gemini N using the Gemini-North Multi-Object Spectrograph (GMOS-N) \citep[][]{Hook2004}. Six NTs were observed with Gemini S using the Gemini-South Multi-Object Spectrograph (GMOS-S) \citep[][]{Gimeno2016} Four NTs were observed with Keck I using the Low Resolution Imaging Spectrometer (LRIS) \citep[][]{Oke1995}. NTs on orbits which have been demonstrated with numerical calculations to be likely temporary captures from the background trans-Neptunian population were not observed \citep[][]{Horner2012, Lin2021}. 

Photometry of NTs was obtained with Sloan Digital Sky Survey (SDSS)-equivalent $g$, $r$, and $i$ filters \citep[][]{Fukugita1996} with the P200, Gemini N, and Gemini S. Observations of NTs with Keck I used the SDSS-equivalent $g$ filter and the, Cousins R and I filters \citep[][]{Cousins1976}. Images were obtained in alternating $g$, $r$/R, and $i$/I sequences to minimize the effect on colour measurements caused by variations in the brightness of the NTs due to their rotation. Exposure times ranged between 30 s and 300 s depending on conditions and the faintness of targets and the number of exposures per filter ranged between 3 and 15. The NTs were tracked at their non-sidereal rates, typically $\approx$0.05\arcsec/s. A complete technical description of the facility and instrumental set for our NT observations is provided in \citep[][]{Bolin2023sup}.

Observations of NTs at all four sites occurred when the targets were as close to opposition as possible at at minimum airmass for maximum throughput and image quality. Seeing varied between 0.8-1.3\arcsec as measured in the WaSP images taken with the P200, between 0.8-1.1\arcsec as measured in the LRIS images taken with Keck I, was $\approx$0.5\arcsec as measured in GMOS-N images taken with Gemini N, and varied between 0.6-0.9\arcsec as measured in GMOS-S images taken with Gemini S. Standard stars from the Panoramic Survey Telescope and Rapid Response System survey \citep[Pan-STARRS][]{Tonry2012} were identified in the same fields as the science observations. A complete list of our observations can be found in Table~S1 of \citet[][]{Bolin2023sup}.

Taking a similar approach as \citet[][]{Bolin20202I}, data from each facility were detrended and flattened using bias frames and flat-field images obtained with an inside-dome flat field panel. Cosmic rays and other blemishes were removed from individual images using the L.A.Cosmic Laplacian cosmic ray identification algorithm \citep[][]{vanDokkum2001}. The data were stacked in each photometric filter to enhance the signal of the NT detections. A complete description of the data reduction for each facility and instrument combination is available in the Supplemental Material \citep[][]{Bolin2023sup}, along with Fig.~S1 showing examples of the P200, Gemini N/S, and Keck I NT detections . 

The photometry of our NT targets and standard stars was performed using an aperture centred on the NT detections with a radius of 1.0-2.5\arcsec that was 1.5-2 times the seeing measured in the images. Sky subtraction was completed by taking the median pixel value within an annulus centred on the NT detection that had an inner radius of 3.0-7.5\arcsec and an outer radius of 6-11\arcsec. The NT photometry obtained with the P200, Keck I, and Gemini N/S was calibrated using the Pan-STARRS photometric catalog \citep[][]{Tonry2012,Chambers2016}.

\section{Results}
The photometric measurements of our 18 NT targets are summarized in Table~1. We have plotted the optical colours of the NTs observed by us and by \citet[][]{Sheppard2006, Sheppard2012, Jewitt2018} and \citet[][]{Lin2019} in $g-i$ vs $g-r$ colour space in Fig.~1. The average $g-i$ value of the 18 NTs is $\approx$0.84, equivalent to a spectral gradient of 8$\%$ / 100 nm normalized to 550 nm and significantly redder than the Sun which has $g-i$ = 0.58 \citep[][]{Haberreiter2017,Willmer2018}. Out of the 18 NTs that we observed, four have $g-i$ colours $\gtrsim$1.2, the rough boundary separating the R and VR groups \citep[][]{Sheppard2012}: 2013 VX$_{30}$, 2011 HM$_{102}$, 2013 TZ$_{187}$, 2015 VV$_{165}$.  Our measurements of the optical colours of the VR NT 2013 VX$_{30}$ with $g-i$ =1.15 $\pm$ 0.17 is broadly consistent with the $g-i$ $\approx$ 1.5 by \citet[][]{Lin2019}, though their $g$-$i$ colour measurement more robustly places it past the 1.2 $g-i$ VR colour boundary.

\begin{table*}
\caption{Photometry.}
\centering
\begin{tabular}{llllll}
\hline
Name            & $\mathrm{m_r^1}$ & $g-r$           & $r-i$           & $g-i$           & S$^2$               \\
                & (mag)          & (mag)         & (mag)         & (mag)         & ($\%$ / 100 nm) \\ \hline
2013 VX$_{30}$  & 23.06  $\pm$ 0.1      & 0.83 $\pm$ 0.19     & 0.32 $\pm$ 0.11     & 1.15 $\pm$ 0.17     & 20.14 $\pm$ 13.18     \\
2012 UD$_{185}$ & 22.28 $\pm$ 0.05     & 0.60 $\pm$ 0.11     & 0.17 $\pm$ 0.06     & 0.77 $\pm$ 0.09     & \: 6.40 $\pm$ 6.32        \\
2014 QO$_{441}$ & 23.35 $\pm$ 0.04 & 0.51 $\pm$ 0.08 & 0.33 $\pm$ 0.06 & 0.84 $\pm$ 0.08 & \:   9.14 $\pm$ 6.09   \\
2012 UV$_{177}$ & 23.01 $\pm$ 0.05 & 0.74 $\pm$ 0.07 & 0.32 $\pm$ 0.07 & 1.06 $\pm$ 0.08 & 16.9 $\pm$ 5.92   \\
2015 VX$_{165}$ & 24.15 $\pm$ 0.10 & 0.42 $\pm$ 0.15     & 0.22 $\pm$ 0.11     & 0.64 $\pm$ 0.12     & \: 1.85 $\pm$ 8.16       \\
2011 HM$_{102}$ & 21.89 $\pm$ 0.05     & 0.91 $\pm$ 0.07     & 0.34 $\pm$ 0.06     & 1.25 $\pm$ 0.06     & 23.90 $\pm$ 4.21       \\
2008 LC$_{18}$  & 23.06 $\pm$ 0.05     & 0.67 $\pm$ 0.08     & 0.21 $\pm$ 0.07     & 0.88 $\pm$ 0.08     & 10.34 $\pm$ 5.79      \\
2013 TZ$_{187}$ & 23.09 $\pm$ 0.04     & 0.81 $\pm$ 0.09     & 0.38 $\pm$ 0.05     & 1.19 $\pm$ 0.08     & 21.79 $\pm$ 6.40       \\
2014 UU$_{240}$ & 22.79 $\pm$ 0.04     & 0.45 $\pm$ 0.07     & 0.07 $\pm$ 0.06     & 0.52 $\pm$ 0.07     & -2.05 $\pm$ 4.74      \\
2015 VW$_{165}$ & 23.02 $\pm$ 0.04     & 0.38 $\pm$ 0.07     & 0.21 $\pm$ 0.05     & 0.59 $\pm$ 0.06     & \: 0.15 $\pm$ 4.05       \\
2014 RO$_{74}$  & 24.01 $\pm$ 0.10 & 0.38 $\pm$ 0.13     & 0.26 $\pm$ 0.15     & 0.64 $\pm$ 0.1      & \: 1.78 $\pm$ 9.09       \\
2014 SC$_{374}$ & 24.05 $\pm$ 0.07 & 0.44 $\pm$ 0.09     & 0.29 $\pm$ 0.11     & 0.73 $\pm$ 0.1      & \: 4.98 $\pm$ 6.98       \\
2013 RL$_{124}$ & 23.76 $\pm$ 0.09 & 0.44 $\pm$ 0.12     & 0.34 $\pm$ 0.14     & 0.78 $\pm$ 0.14     & \: 6.55 $\pm$ 9.52       \\
2015 VV$_{165}$ & 23.51 $\pm$ 0.08 & 0.93 $\pm$ 0.11     & 0.46 $\pm$ 0.11     & 1.39 $\pm$ 0.12     & 29.68 $\pm$ 9.05      \\
2014 YB$_{92}$  & 23.44 $\pm$ 0.06     & 0.42 $\pm$ 0.08     & 0.15 $\pm$ 0.09     & 0.57 $\pm$ 0.09     & -0.66 $\pm$ 5.92      \\
2013 TK$_{227}$ & 23.79 $\pm$ 0.06     & 0.63 $\pm$ 0.08     & 0.38 $\pm$ 0.12     & 1.01 $\pm$ 0.12     & 14.67 $\pm$ 9.01       \\
2013 RC$_{158}$ & 23.54 $\pm$ 0.07     & 0.32 $\pm$ 0.09     & 0.27 $\pm$ 0.08     & 0.59 $\pm$ 0.08     & \: 0.25 $\pm$ 5.34       \\
2015 VU$_{207}$ & 21.98 $\pm$ 0.04     & 0.58 $\pm$ 0.07     & 0.09 $\pm$ 0.05     & 0.67 $\pm$ 0.07     & \: 2.66 $\pm$ 4.61      \\\hline
Solar colours$^3$ &      & 0.46  $\pm$ 0.01   & 0.12 $\pm$ 0.01    & 0.58  $\pm$ 0.01   &      \\\hline
\end{tabular}
\begin{tablenotes}
\item \textbf{Notes.} (1) Apparent $r$-band magnitude, (2) spectral gradient using the $g$ and $i$ measurements normalized to 550 nm, (3) from \citet[][]{Haberreiter2017} and \citet[][]{Willmer2018}.
\end{tablenotes}
\end{table*}

\begin{figure}\centering
\hspace{0 mm}
\centering
\includegraphics[width=1\linewidth]{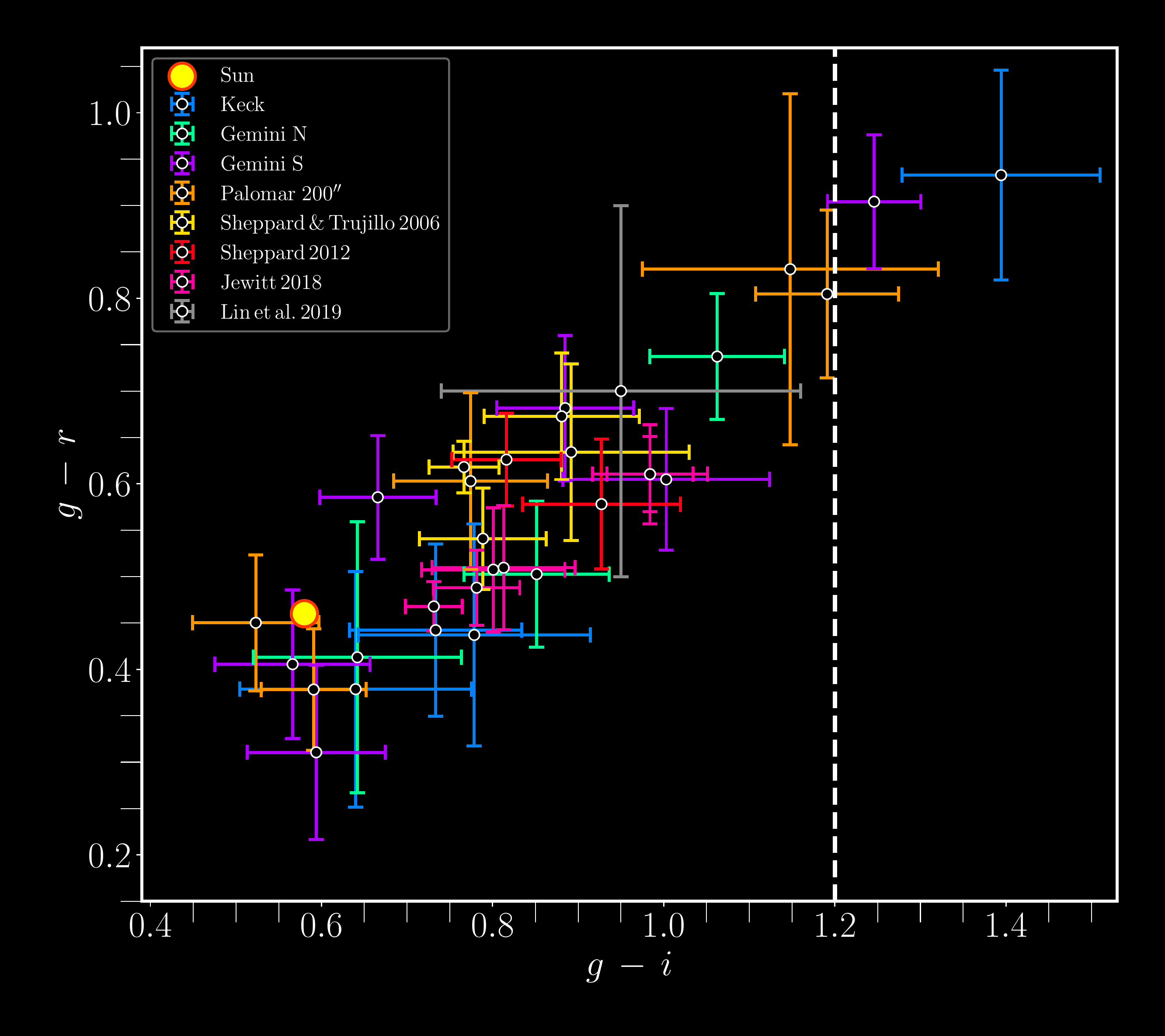}
\caption{Sloan $g$, $r$, and $i$ colours for the NTs observe in this work. Sloan NT colours from \citet[][]{Sheppard2006, Sheppard2012, Jewitt2018} and \citet[][]{Lin2019} are also plotted. The vertical dashed line at $g-i$ = 1.2 indicates the rough dividing line between the R and VR colour groups.The colours of the sun are plotted at $g-i$ = 0.58 and $g-r$ = 0.46 \citep[][]{Haberreiter2017,Willmer2018}.}
\end{figure}

One of the NTs we observed, 2011 HM$_{102}$, has $g-r$ = 0.91 $\pm$ 0.07 and $r-i$ = 0.34 $\pm$ 0.06 placing it into the VR category with $g-i$ = 1.25 $\pm$ 0.06. \citet[][]{Parker2013} observed 2011 HM$_{102}$ and found $r-i$ = 0.31 $\pm$ 0.04, consistent with our measurements, but found $g-r$ = 0.51 $\pm$ 0.04, significantly bluer than our measured  $g-r$ = 0.91 $\pm$ 0.07. The difference could be due to underestimated uncertainties in the $g-r$ colour measurement by ourselves or by \citet[][]{Parker2013}, or due to lightcurve variations. The SNR of our composite 2011 HM$_{102}$ $g$ and $r$ observations was $\sim$20 each and is consistent with the SNR expected from integration time calculations simulating the conditions of our observations\footnote{\url{https://www.gemini.edu/instrumentation/gmos/exposure-time-estimation}}. To test the latter hypothesis, we measured the $g$ magnitude of 2011 HM$_{102}$ in the $g$ images taken at Gemini S taken in the $g$, $r$, and $i$ sequence on 2022 Jul 18 UTC and found that there were no significant variations in the brightness of 2011 HM$_{102}$ in excess of $\approx$0.05 over the roughly half-hour observing sequence. In addition, our $g-i$ measurements of 2014 QO$_{441}$ of $g-i$ = 0.84 $\pm$ 0.08 and of 2014 UU$_{240}$ of $g-i$ = 0.52 $\pm$ 0.07 are generally consistent with the values measured by \citet[][]{Lin2019}.

\section{Discussion and conclusion}
An initial impression of the NT colours in Fig.~1 is an apparent lack of a bimodal colour distribution. Following the example of \citet[][]{Jewitt2018}, we have compared the $g-i$ colours of NTs with those of other dynamical classes. Using the optical colours compiled by the Minor Bodies in the Outer Solar System (MBOSS) database \citep[][]{Hainaut2012}, we have plotted the cumulative $g-i$ colour distribution of the Jovian Trojans (JTs), Centaurs, Scattered Disc objects, Plutinos, Resonant objects, HCs, CCs and Detached Objects \citep[][]{Gladman2008} with the colours of NTs in Fig.~2. By visual inspection, the cumulative $g-i$ distribution of the JTs is distinct in lacking any VR objects compared to objects of the other TNO dynamical classes. However, it must be noted that while they lack VR objects, the Jupiter Trojans are bimodal in colour, albeit with bluer mean colours compared to the TNO population \citep[e.g.,][]{Wong2014trojan,Wong2015trojan}. The cumulative distribution of the $g-i$ colours of the NTs is located between these two groups, containing more VR objects than JTs, but disproportionately fewer VR objects compared to the other TNO classes, especially the CCs.

\begin{figure}
\centering
\includegraphics[width=1\linewidth]{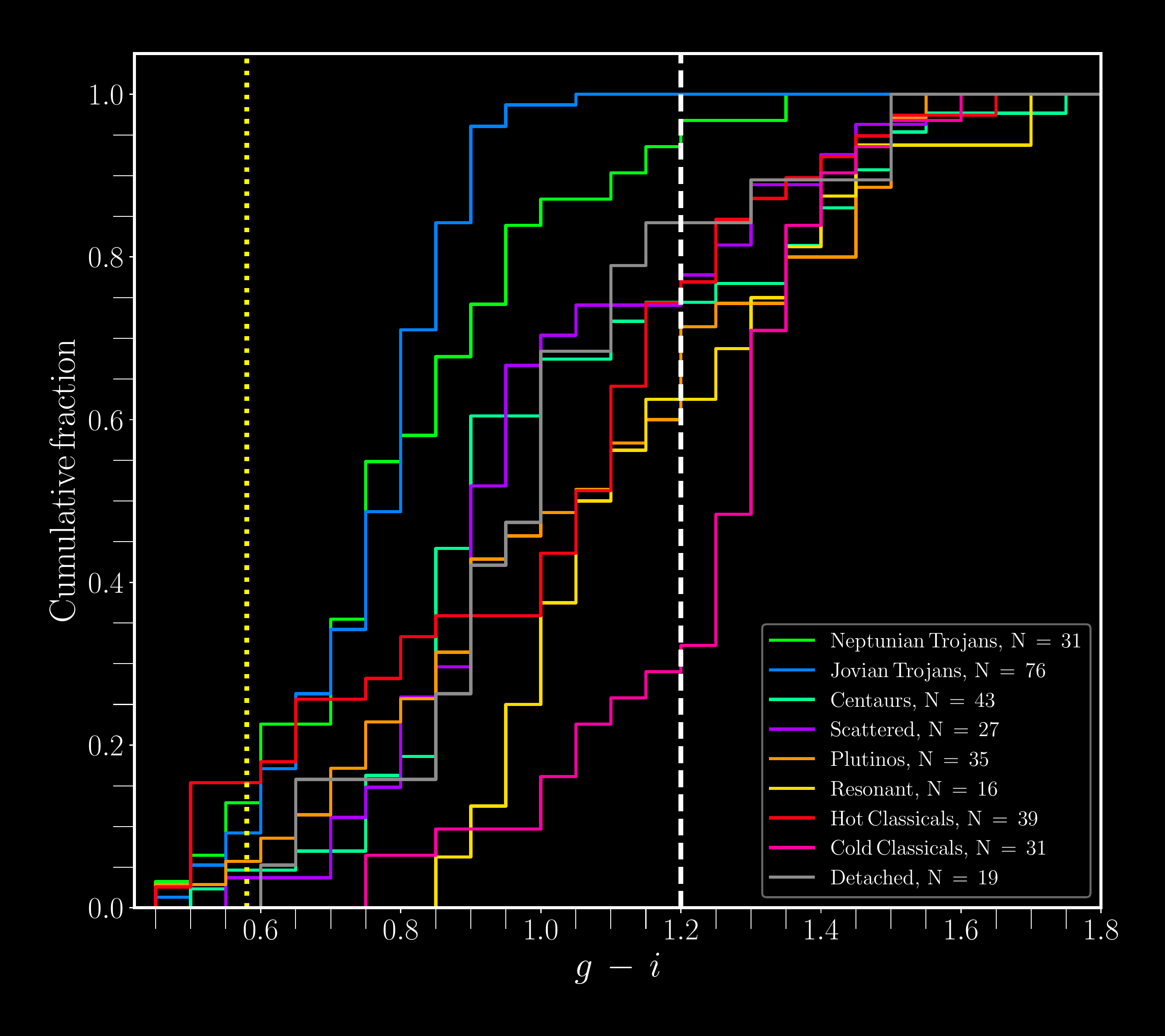}
\caption{Cumulative distribution of the $g-i$ colours for the NTs and the $g-i$ colours for Jupiter Trojans, Centaurs, Scattered Disc Objects, Plutinos, Resonant objects, HCs, CCs and Detached objects taken from \citet[][]{Hainaut2012}. The white vertical dashed line at $g-i$ = 1.2 indicates the rough division line between R and VR objects. The yellow vertical dotted line indicates the colour of the Sun at $g-i$ = 0.58 \citep[][]{Haberreiter2017,Willmer2018}.}
\end{figure}

To quantify the differences between the $g-i$ distribution of the NTs and the $g-i$ distribution of other dynamical classes, we apply the Kolmogorov-Smirnov (KS) test which measures the maximum difference between two cumulative distributions \citep[][]{Darling1957}. The KS method tests the null hypothesis that two cumulative distributions are drawn from the same parent distribution. We have also applied the Kuiper variant of the KS test which is more sensitive to differences between distributions at their edges \citep[][]{Kuiper1960TestsCR}. Table~2 shows the associated statistical score and p-value of the KS and Kuiper tests between the NTs and other TNO dynamical classes. The statistical score is a quantified measure of the maximum difference between two cumulative distributions with a larger statistical score corresponding to a lower p-value for datasets of similar size. The comparison between the NTs and the CCs results in the largest statistical score of 0.74 for both the KS and Kuiper test corresponding to a p-value of $<$ 0.0001. The comparison between the NTs and JTs results in the smallest statistical score of 0.22 corresponding to a p-value of 0.2138 for the KS test and a statistical score of 0.28 and a p-value of 0.0882 for the Kuiper test. The KS and Kuiper tests between the NTs and the Centaurs, Scattered Disc objects, Plutinos, Resonant objects, HCs and Detached objects have p-values $<$0.005.

\begin{table}
\caption{Kolmogorov-Smirnov (KS) and Kuiper (K) variant NT statistical score and p-values (P).}
\begin{tabular}{llllll}
\hline
Class             & N & KS Score & KS P & K Score & K P \\
NTs & 31     & 0.0       & 1.0        & 0.0          & 1.0            \\
JTs    & 76     & 0.2186    & 0.2138     & 0.2830       & 0.0882         \\
Centaur           & 43     & 0.4449    & 0.0010     & 0.3946       & 0.0012         \\
Scattered Disc    & 27     & 0.4182    & 0.0086     & 0.4053       & 0.0008         \\
Plutinos          & 35     & 0.4203    & 0.0039     & 0.4389       & $<$0.0001      \\
Resonant          & 16     & 0.7137    & $<$0.0001  & 0.6992       & $<$0.0001      \\
HCs    & 39     & 0.4797    & 0.0004     & 0.5833       & $<$0.0001      \\
CCs   & 31     & 0.7419    & $<$0.0001  & 0.7419       & $<$0.0001      \\
Detached objects  & 19     & 0.5195    & 0.0019     & 0.4873       & $<$0.0001  \\ \hline   
\end{tabular}
\end{table}

The $g-i$ colour distribution of the NTs is distinct compared to other TNO classes. Previous studies show a much larger p-value for the tests between the cumulative optical colour distribution of the NTs and JTs \citep[][]{Jewitt2018}. Our expanded sample with additional VR NTs implies dissimilarity in the $g-i$ colours of NTs and JTs, although it only rules out the null hypothesis at the 1-2-$\sigma$ level with a p-value of $\approx$0.08-0.21. The small number of the VR NTs and the large error bars on the $g-i$ colours may make drawing a strong conclusion about the differences between the optical colours of NTs and JTs difficult.

TNO evolutionary models predict that the observed proportion of VR and R objects in different TNO classes is a result of the separation between VR and R objects in the original TND located at a radial distance from the Sun denoted as $r_*$ \citep[][]{Nesvorny2020}. The combined sample of our observed NTs with those from the literature results in a VR to R ratio of $\approx$ 1:8. Although the location of $r_*$ may also be affected by the density profile of the original TND, a higher proportion of VR objects to R objects may imply a closer in value of $r_*$ compared to a lower proportion. In the case of a disc with an exponential density profile, an NT VR to R ratio of 1:8 may imply a $r_*$ interior to 35 au whereas a truncated profile may imply a $r_*$ interior to 30 au. In either case, the discovery of additional NTs and measurements of their optical colours will provide additional constraints on the compositional gradient of the original TND. In addition to the location of the transition boundary between R and VR objects in the original TBD, the colours of TNOs could also be affected by post formation evolutionary effects such as collisions and thermal processing \citep[][]{McKinnon2008}.

\section*{Acknowledgements}
The authors appreciate the help from O. Oberdorf with the reduction of GMOS-N and GMOS-S images.  C.F.~acknowledges support from the Heising-Simons Foundation (grant $\#$2018-0907). We wish to recognize and acknowledge the cultural role and reverence that the summit of Maunakea has always had within the indigenous Hawaiian community. The authors wish to recognize and acknowledge the cultural significance that Palomar Mountain has for the Pauma Band of the Luise\~{n}o Indians. Based on observations obtained at the international Gemini Observatory, a program of NSF's NOIRLab, which is managed by the Association of Universities for Research in Astronomy (AURA) under a cooperative agreement with the National Science Foundation on behalf of the Gemini Observatory partnership. Some of the data presented herein were obtained at the W. M. Keck Observatory, which is operated as a scientific partnership among the California Institute of Technology, the University of California and the National Aeronautics and Space Administration.%%%%%%%%%%%%%%%%%%%%%%%%%%%%%%%%%%%%%%%%%%%%%%%%%%
\section*{Data Availability}
The data underlying this article will be shared on reasonable request to the corresponding author.
\section*{Supplemental Material}
The supplemental material for this manuscript is available online.



\bibliographystyle{mnras}
\bibliography{/Users/bolin/Dropbox/Projects/NEOZTF_NEOs/neobib} % if your bibtex file is called example.bib

%\begin{thebibliography}{99}
%\end{thebibliography}

%%%%%%%%%%%%%%%%%%%%%%%%%%%%%%%%%%%%%%%%%%%%%%%%%%

%%%%%%%%%%%%%%%%% APPENDICES %%%%%%%%%%%%%%%%%%%%%

\renewcommand{\thefigure}{S\arabic{figure}}
\setcounter{figure}{0}
\renewcommand{\thetable}{S\arabic{table}}
\renewcommand{\theequation}{S\arabic{equation}}
\renewcommand{\thesection}{S\arabic{section}}
\setcounter{section}{0}
\cleardoublepage
\setcounter{page}{1}
\renewcommand\thepage{S\arabic{page}}
\section*{Supplemental Material}
\appendix
\renewcommand{\thefigure}{S\arabic{figure}}
\setcounter{figure}{0}
\renewcommand{\thetable}{S\arabic{table}}
\renewcommand{\theequation}{S\arabic{equation}}
\renewcommand{\thesection}{S}
\setcounter{section}{0}
\subsection{Observational details}
We obtained optical $g$, $r$/R, and $i$/I photometry of 18 NTs with the Hale 5.1 m telescope (P200 hereafter) at Palomar Observatory, the Gemini North 8.1 diameter telescope and the Keck I 10 m diameter telescope at Maunakea Observatory, and the Gemini South 8.1m diameter telescope at Cerro Pach\'{o}n. A record of our observations is available in Table~S1. Examples of NT detections from each facility is available in Fig.~S1.

\noindent\textit{Palomar Hale 5.1 m diameter telescope, P200:} The Wafer-Scale Imager for Prime (WaSP) instrument mounted at prime focus on the P200 was used to observe 2013 VX$_{30}$ and 2012 UD$_{185}$ on 2020 Dec 22 UTC under program 2020B-P27 (PI: B. Bolin), and 2013 TZ$_{187}$, 2014 UU$_{240}$, and 2015 VW$_{165}$ on 2021 October 30 UTC under program 2021B-P17 (PI: B. Bolin). The WaSP detector array consists of a 6144 $\times$ 6160 Teledyne e2v array with a pixel scale of 0.19\arcsec pixel$^{-1}$ \citep[][]{Nikzad2017}. The NTs were observed with SDSS $g$ ($\mathrm{\lambda_{eff}}$ = 467.2 nm, FWHM = 126.3 nm), $r$ ($\mathrm{\lambda_{eff}}$ = 614.1 nm, FWHM = 115.0 nm), and $i$ ($\mathrm{\lambda_{eff}}$ = 745.8 nm, FWHM = 123.9 nm) filters 
\citep[][]{Fukugita1996} and were rotated to minimize the effects of rotational brightness variations on the $g$, $r$, and $i$ colour measurements. The telescope was tracked at the on-sky rate of motion of the NT targets.
\\
\textit{Gemini North 8.1 m diameter telescope, Gemini N:} The Gemini-North Multi-Object Spectrograph (GMOS-N) on the Gemini N telescope was used to observe 2014 QO$_{441}$ and 2014 UV$_{177}$ on 2021 Feb 1 UTC, and 2015 VX$_{165}$ on 2022 Feb 17 UTC under program GN-2021A-FT-201 (PI: B. Bolin). The GMOS-N detector array consists of three 2048 $\times$ 4176 Hamamatsu chips separated by 67-pixel gaps with a pixel scale of 0.08\arcsec pixel$^{-1}$ \citep[][]{Hook2004}. The NTs were observed with SDSS-equivalent $g$, $r$, and $i$ filters and were swapped to minimize the effects of rotational brightness variations on the $g$, $r$, and $i$ colour measurements. The conversions from \citep[][]{Schwamb2019} were used to transform the GMOS-N $g$, $r$, and $i$ magnitudes to SDSS $g$, $r$, and $i$ magnitudes. Exposure times between 100 s and 233 s were used depending on the brightness of the targets and the filters being used. The telescope was tracked at the on-sky rate of motion of the NT targets.
\\
\textit{Gemini South 8.1 m diameter telescope, Gemini S:} The Gemini-South Multi-Object Spectrograph (GMOS-S) on the Gemini S telescope was used to observe 2014 RO$_{74}$, 2014 SC$_{374}$, and 2013 RL$_{124}$ on 2022 July 18 UTC under program GS-2021B-Q-318 (PI: B. Bolin), and 2014 YB$_{92}$ and 2013 TK$_{227}$ on 2022 Nov 28 UTC, 2013 RC$_{158}$ on 2022 Nov 29 UTC, and 2015 VU$_{207}$ on 2022 Dec 10 UTC under program GS-2022B-FT-109 (PI: B. Bolin). The GMOS-S detector array consists of three 2048 $\times$ 4176 Hamamatsu chips separated by 61-pixel gaps with a pixel scale of 0.08\arcsec pixel$^{-1}$ \citep[][]{Gimeno2016}. SDSS-equivalent $g$, $r$, and $i$ filters were used to observe the NTs and were cycled to reduce the effects of brightness variations caused by the rotation of the objects on the colour measurements. The GMOS-S $g$, $r$, and $i$ photometric measurements were converted to SDSS equivalent $g$, $r$, and $i$ measurements using the conversion formulae in \citep[][]{Schwamb2019}. Exposure times between 30 s and 170 s were used depending on the brightness of the targets and the filters being used. The telescope was tracked at the on-sky rate of motion of the NT targets.
\\
\textit{Keck I Telescope:} The Low Resolution Imaging Spectrometer (LRIS) \citep[][]{Oke1995} on the Keck I telescope was used to observe 2014 RO$_{74}$, 2014 SC$_{374}$, and 2013 RL$_{124}$ on 2022 February 3 UTC under program 2022A-C277, and 2015 VV$_{165}$ on 2022 March 3 UTC under program 2022A-C244 (PI: J. van Roestel). LRIS includes two separate blue and red camera channels separated by a dichroic. The blue camera consists of two 2k $\times$ 4k Marconi CCD arrays and the red camera consists of two science grade Lawrence Berkeley National Laboratory 2k $\times$ 4k CCD arrays. Both cameras have a pixel scale of 0.135 arcsec pixel$^{-1}$. The 460 nm dichroic was used in combination with an SDSS-equivalent $g$ filter in the blue camera and Cousins R ($\mathrm{\lambda_{eff}}$ = 649.2 nm, FWHM = 167.1 nm) and I ($\mathrm{\lambda_{eff}}$ = 799.3 nm, FWHM = 152.3 nm) filters \citep[][]{Cousins1976} in the red camera. Similar observational strategies described in \citet[][]{Bolin2020CD3, Bolin2021LD2,Bolin2022AV2} were applied to the observation of the NTs. The telescope was tracked at the on-sky rate of motion of the NT targets. Exposure times of 210 s were used on the observations taken on 2022 February 3 UTC and Exposure times of 300 s were used on observations taken on 2022 March 3 UTC. Exposures were taken in the $g$ filter using the blue camera simultaneously with the R and I filter exposures taken with the red camera. The Cousins R and I photometry measurements were transformed to SDSS $r$ and $i$ equivalent brightnesses were performed using the conversions in \citet[][]{Jordi2006}.
\\
\begin{table*}
\caption{Observational details.}
\centering
\begin{tabular}{llllllllllll}
\hline
Name            & Facility & Instrument & UT Date    & $\Delta^1$ & r$_H^2$  & $\alpha^3$     & Seeing$^4$                   & Airmass & Exp. $g$ & Exp. $r$/R$^5$ & Exp. $i$/I$^6$ \\
                &          &            &            & (au)     & (au)   & ($^{\circ}$) & (\arcsec) &         &        &          &         \\ \hline
2013 VX$_{30}$  & P200     & WaSP       & 2020 Dec 22 & 26.897   & 27.573 & 1.5          & 1.3                      & 1.3     & 1800   & 900      & 900     \\
2012 UD$_{185}$ & P200     & WaSP       & 2020 Dec 22 & 30.716   & 31.525 & 1.0          & 1.1                      & 1.6     & 900    & 900      & 900     \\
2014 QO$_{441}$ & Gemini N & GMOS-N     & 2021 Feb 1 & 33.280   & 33.274 & 1.7          & 0.5                      & 1.2     & 300    & 300      & 300     \\
2012 UV$_{177}$ & Gemini N & GMOS-N     & 2021 Feb 1 & 28.702   & 28.840 & 1.9          & 0.5                      & 1.2     & 699    & 300      & 300     \\
2015 VX$_{165}$ & Gemini N & GMOS-N     & 2021 Feb 17 & 32.253   & 32.307 & 1.8          & 0.5                      & 1.2     & 1200   & 600      & 600     \\
2011 HM$_{102}$ & Gemini S & GMOS-S     & 2021 Jul 18 & 27.400   & 28.399 & 0.4          & 0.9                      & 1.0     & 360    & 90       & 90      \\
2008 LC$_{18}$  & Gemini S & GMOS-S     & 2021 Jul 18 & 31.393   & 32.382 & 0.4          & 0.8                      & 1.0     & 1200   & 360      & 360     \\
2013 TZ$_{187}$ & P200     & WaSP       & 2021 Oct 30 & 29.854   & 30.822 & 0.4          & 0.9                      & 1.3     & 2100   & 1200     & 1800    \\
2014 UU$_{240}$ & P200     & WaSP       & 2021 Oct 30 & 29.656   & 30.556 & 0.8          & 0.8                      & 1.5     & 600    & 600      & 600     \\
2015 VW$_{165}$ & P200     & WaSP       & 2021 Oct 30 & 28.945   & 29.847 & 0.8          & 0.8                      & 1.3     & 2100   & 1200     & 2100    \\
2014 RO$_{74}$  & Keck I   & LRIS       & 2022 Feb 3 & 31.194   & 31.193 & 1.8          & 0.9                      & 1.3     & 630    & 420      & 210     \\
2014 SC$_{374}$ & Keck I   & LRIS       & 2022 Feb 3 & 33.057   & 33.038 & 1.7          & 0.8                      & 1.3     & 420    & 420      & 420     \\
2013 RL$_{124}$ & Keck I   & LRIS       & 2022 Feb 3 & 29.476   & 29.371 & 1.9          & 1.1                      & 1.3     & 630    & 420      & 210     \\
2015 VV$_{165}$ & Keck I   & LRIS       & 2022 Mar 7 & 28.481   & 28.242 & 1.9          & 0.9                      & 1.2     & 600    & 300      & 300     \\
2014 YB$_{92}$       & Gemini S & GMOS-S     & 2022 Nov 28 & 29.880   & 30.710 & 1.0          & 0.8                      & 1.3     & 435    & 138      & 138     \\
2013 TK$_{227}$      & Gemini S & GMOS-S     & 2022 Nov 28 & 30.688   & 31.577 & 0.8          & 0.7                      & 1.3     & 2550   & 700      & 550     \\
2013 RC$_{158}$      & Gemini S & GMOS-S     & 2022 Nov 29 & 30.318   & 31.159 & 1.0          & 0.7                      & 1.4     & 840    & 285      & 285     \\
2015 VU$_{207}$      & Gemini S & GMOS-S     & 2022 Dec 10 & 29.952   & 30.548 & 1.5          & 0.6                      & 1.0     & 300    & 90       & 90      \\ \hline
\end{tabular}
\begin{tablenotes}
\item \textbf{Notes.} (1) Geocentric distance, (2) heliocentric distance, (3) phase angle, (4) measured in science images, (5) $r$-band images were taken with the P200/WaSP, Gemini N/GMOS-N and Gemini S/GMOS-S. R-band images were taken with Keck/LRIS, (6) $i$-band images were taken with the P200/WaSP, Gemini N/GMOS-N, and Gemini S/GMOS-S. I-band images were taken with Keck/LRIS.
\end{tablenotes}
\end{table*}

\begin{figure}
\centering 
\includegraphics[width=1\linewidth]{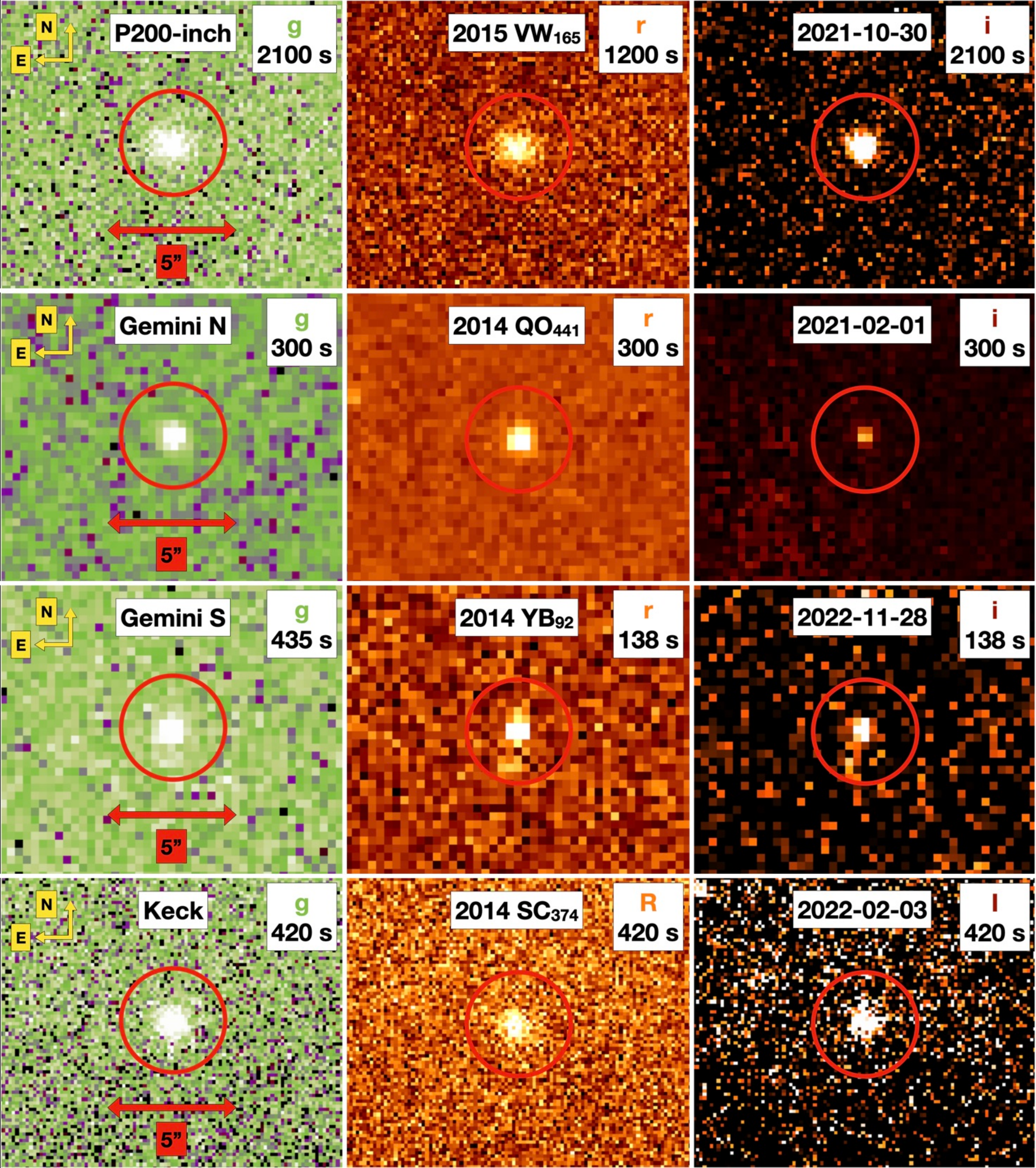}
\caption{Postage stamps of NT detections circled in red obtained with the P200, Gemini-N, Gemini-S, and Keck I telescopes. The images were aligned and stacked according to the position and motion of the target. The left column consists of detections made in $g$-band. The centre column consists of detections made in $r$-band for observations with the P200, Gemini N, and Gemini S telescopes and R-band for observations with the Keck I telescope. The right column contains detections made in $i$-band for observations with the P200, Gemini N, and Gemini S telescopes and I-band for observations with the Keck I telescope. The cardinal directions, image scale, and total integration times are indicated.}
\end{figure}

%%%%%%%%%%%%%%%%%%%%%%%%%%%%%%%%%%%%%%%%%%%%%%%%%%

% Don't change these lines
\bsp	% typesetting comment
\label{lastpage}
\end{document}